\documentclass[12pt]{article}
\begin{document}
\title{The elastoplastic response of and moisture diffusion 
through a vinyl ester resin--clay nanocomposite}

\author{A.D. Drozdov, J. deC. Christiansen\\
Department of Production\\ 
Aalborg University\\
Fibigerstraede 16\\
DK--9220 Aalborg, Denmark\\
and\\
R.K. Gupta, A.P. Shah\footnote{Present address: Triton Systems Inc., 
200 Turnpike Road, Chelmsford, MA 01824, USA}\\
Department of Chemical Engineering and Constructed Facilities Center\\
West Virginia University\\ 
P.O. Box 6102\\ 
Morgantown, WV 26506, USA}
\date{}
\maketitle

\begin{abstract}
Experimental data are reported on the elastoplastic response of
and moisture diffusion through a vinyl ester resin--montmorillonite 
clay nanocomposite with various amounts of filler.
Two simple models are developed for the elastoplastic behavior
of a nanocomposite and for the anomalous diffusion of penetrant
molecules.
Adjustable parameters in the constitutive equations are found by
fitting the observations.
It is revealed that some critical concentration of filler exists (about 1 wt.-\%):
in the sub-critical region of concentrations, molecular mobility of
the host polymer strongly decreases with the clay content,
whereas in the post-critical domain, the filler fraction weakly
affects mobility of chains.
\end{abstract}
\vspace*{10 mm}

\newpage

\section{Introduction}

This paper is concerned with the elastoplastic behavior of and
moisture diffusion through a nanocomposite consisting of
a vinyl ester resin matrix filled with montmorillonite clay particles.

The choice of vinyl ester resin is explained by the fact that this
thermosetting polymer is widely used as a matrix for glass-reinforced
polymer composites employed for construction and repair of bridges and
other civil structures (Valea et al., 1998; Patel et al., 1999).

Montmorillonite (MMT) is a commonly used clay with a layered structure
that is constructed of two tetrahedral sheets of silica surrounding an
octahedral sheet of alumina or magnesia (Masenelli-Varlot et al., 2002).
The layers (with thickness of 1 nm) are stacked by weak dipole forces,
while the galleries between the layers are occupied by metal cations.
Processing of clay particles for preparation of an intercalated
nanocomposite consists of two stages: 
\begin{enumerate}
\item
The growth of the basal (interlayer) spacing by exchanging the metal cations
with an intercalating reagent (conventionally, alkylammonium ions).
This process facilitates intercalation of polymer chains into the galleries
between the clay layers (Chang et al., 2002).

\item
Treatment of intercalated particles by a compatibilizer (a swelling agent,
a functional oligomer, or a low molecular weight polymer) 
to improve miscibility between 
the organics-modified silicate layers and the host matrix
(Tyan et al., 2000; Ma et al., 2001; Chang et al., 2002).
\end{enumerate}
Preparation and analysis of micro-structure and physical properties
of nanocomposites with polymeric matrices filled with MMT  
clay have been a focus of attention in the past couple of years.
It was found that polymer--clay nanocomposites possess many desirable
properties, such as increased moduli and strength,
reduced gas permeabilities,
decreased adsorption of organic liquids, etc. (Kim et al., 2002).
However, no detailed analysis of interaction between molecules of 
a host matrix and the silica sheets has been performed so far.

The objective of the present study is to investigate how the presence
of intercalated MMT clay affects the mobility of polymeric chains.
We focus on two main processes, where this mobility is revealed: 
\begin{enumerate}
\item
elastoplasticity (which is attributed to sliding of
junctions in the polymeric network with respect to their reference
positions in the bulk material),

\item
moisture diffusion (whose rate is governed by the ability of penetrant 
molecules to move through the host network).
\end{enumerate}
In civil engineering applications, these two processes are of essential
importance for the prediction of durability and performance of constructions.

To evaluate the effect of nano-particles on the mobility of macromolecules,
we focus on a model nanocomposite where no compatibilizer is used.
Because of low miscibility between the clay particles and the vinyl ester resin,
we do not expect that the presence of intercalated clay particles noticeably 
improves the mechanical and transport properties of a nanocomposite compared
to those of the neat resin (Ma et al., 2001).
Our aim is to demonstrate that the concentration of nano-particles affects
the parameters that characterize mobility of polymeric chains in mechanical 
and diffusion tests in similar ways.

In the past two decades, a number of constitutive models have been
proposed for the elastoplastic and viscoplastic behavior of glassy polymers,
see, e.g., Boyce et al. (1988), Mangion et al. (1992),
Hasan and Boyce (1995), David et al. (1997), Drozdov (2001)
and the references therein.
Three shortcomings of these models should be mentioned:
\begin{itemize}
\item
the stress--strain relations were developed to correctly describe
the mechanical behavior of polymers near the yield point (which is 
not the case for vinyl ester resin where fracture occurs earlier 
than yielding, see Figures 1 to 5 below),

\item
the constitutive equations contain a number of adjustable parameters
which cannot be determined with a high level of accuracy by fitting data
in uniaxial tensile tests,

\item
the material constants are not explicitly expressed in terms of the physical
quantities that describe molecular mobility.
\end{itemize}
To assess constrains imposed by the clay particles on the mobility
of macromolecules in a host matrix, we develop a simple model
that contains only three adjustable parameters with a transparent
physical meaning.
Only one of these parameters reflects mobility of chains, and it can
be easily found by fitting stress--strain curves in uniaxial tensile tests
with small strains.

Diffusion of moisture in polymer composites has attracted substantial
attention in the past decade, see, e.g., 
Cai and Weitsman (1994),
Bavisi et al. (1996),
Vanlandingham et al. (1999),
Marcovich et al. (1999),
Roy et al. (2000),
Chen et al. (2001),
Roy et al. (2001),
Uschitsky and Suhir (2001),
to mention a few.
It was demonstrated that diffusion of penetrants in several glassy polymers 
is anomalous (non-Fickian), with the mass gain proportional
to some power of time that differs from 0.5 (the latter is characteristic for the 
Fickian diffusion).

Two approaches were recently proposed to model the anomalous diffusion.
According to the first (the fractional dynamics concept),
conventional derivatives with respect to time and spatial variables
in the diffusion equation are replaced by fractional derivatives,
see Metzler and Klafter (2000) and the references therein.
According to the other (Roy at al., 2000), a constant coefficient 
of diffusion is replaced by a decreasing function of time (by analogy 
with a relaxation modulus for a viscoelastic solid).

Our experimental data (see Figures 6 to 10 below) reveal that moisture
diffusion in the neat vinyl ester resin is practically Fickian, while the
anomalous transport of water molecules is observed with an increase in the
clay concentration.
This phenomenon may be explained by the fact that the water molecules are
bounded to hydrophilic surfaces of clay layers, 
where they become immobilized.
A two-stage diffusion process with immobilization of penetrant molecules
has been previously studied by Carter and Kibler (1978)
and Gurtin and Yatomi (1979).
Unlike these works, we assume that bounded molecules cannot leave the sites
where they were immobilized.
This allows the number of adjustable parameters to be substantially reduced.
The diffusion process is governed by differential equations with
only three material constants that are found by matching the water uptake curves 
plotted in Figures 6 to 10.

The exposition is organized as follows.
Observations in uniaxial tensile tests and in moisture diffusion tests
are reported in Section 2.
Constitutive equations for the elastoplastic response are derived in
Section 3 by using the laws of thermodynamics.
Adjustable parameters in these relations are determined in Section 4.
A model for moisture diffusion is developed in Section 5
and its material constants are found in Section 6.
A brief discussion of our findings is presented in Section 7.
Some concluding remarks are formulated in Section 8.

\section{Experimental procedure}

\subsection{Preparation of samples}

The polymer used was DERAKANE 411-350 epoxy vinyl ester 
resin (Dow Chemical Co.) containing 45 wt.-\% of dissolved styrene. 
To cure the resin at room temperature, it was mixed with 0.5 wt.-\% of 6\% cobalt 
naphthenate catalyst (Sigma Aldrich Co.). 
Additionally, 0.05\% of 99\% N,N dimethyl aniline 
(Lancaster Synthesis, Pelham, NH, USA) was used as 
an accelerator, and 1.5\% of methyl ethyl ketone peroxide 
with 9\% of active oxygen (Sigma Aldrich Co.) as an initiator. 

A montmorillonite clay Cloisite 10A (Southern Clay Products Inc.,
Gonzales, TX, USA) treated with benzyl (hydrogenated tallow alkyl) 
dimethyl quaternary ammonium chloride was used as received.
Neat resin coupons were cast by pouring 
the reaction mixture into Teflon molds with the dimensions
50 $\times$ 12.5 mm and the thickness ranging from 0.2 to 0.6 mm. 
The organically treated clay was added to the liquid resin and manually stirred. 
The mixture was then degassed in a vacuum oven to remove air bubbles. 
Afterwards, the catalyst, initiator and accelerator were added.
The mixture was allowed to cure at room temperature for 24 hours, 
and it was subsequently post cured in an oven for 3 hours at 90 $^{\circ}$C. 

TEM (transmission electron microscopy) micrographs 
of the nanocomposite samples show that the clay is 
reasonably uniformly distributed, 
and the silica sheets are randomly oriented (Shah et al., 2002).
The clay exists as expanded aggregates made up of 2 to 10 sheets
with the interlayer distance ranging from 4.4 to 5.0 nm,
which confirms the formation of an intercalated nanocomposite.

\subsection{Mechanical tests}

Dumbbell specimens for mechanical tests were cut following 
ASTM D638 specification. 
Tensile tests were performed at ambient temperature by
using a 100 kN Instron machine model 8501 with a cross-head 
speed of 0.254 mm/min. 
This strain rate ensures practically isothermal loading conditions.
The strain was measured independently using a strain gauge 
affixed to the mid-point of the specimen. 
The tensile force was measured by a standard load cell.
The engineering stress, $\sigma$, was determined as the ratio
of the tensile force to the cross-sectional area of a specimen
in the stress-free state.
For any concentration of filler in a nanocomposite, $\nu$,
four specimens were tested.
Typical stress--strain diagrams are presented in Figures 1 to 5.
These figures demonstrate that the response of the neat resin
is strongly nonlinear, but the nonlinearity of the stress--strain
curves is reduced with an increase in the filler content.

DSC (differential scanning calorimetry) measurements
show that the glass transition temperature of the neat vinyl
ester resin is about 97 $^{\circ}$C, and it increases with
concentration of clay reaching 117 $^{\circ}$C at $\nu=5.0$ wt.-\%
(Shah et al., 2002).
Because the material at room temperature is far below 
its glass transition point, we disregard its viscoelastic
response and attribute the nonlinearity of the stress--strain
curves to the elastoplastic behavior.

\subsection{Diffusion tests}

Diffusion tests were performed by immersing samples
with rectangular cross-section  
in distilled water at room temperature ($T=25$ $^{\circ}$C). 
The samples had a dry mass ranging from 120 to 400 mg.
They were stored in a controlled humidity chamber, 
and contained 0.05 $\pm$ 0.005 wt.-\% water 
at the beginning of diffusion tests.
The samples were periodically removed, 
blotted dry with a lint-free tissue, 
weighed and re-immersed in the water. 
A typical experiment lasted ten days, 
and, on the first day, readings were taken as frequently 
as every 15 min. 
The balance used had an accuracy of 1 mg, 
and 3 to 5 replicate runs were carried out 
for a given set of condition.
Typical dependences of the relative moisture uptake, $\Phi$,
versus the reduced time, $\bar{t}$, are plotted in Figures 6 to 10.
The reduced time, $\bar{t}$, is given by
\begin{equation}
\bar{t}=\frac{t}{4l^{2}},
\end{equation}
where $t$ is time elapsed from the beginning of a test,
and $2l$ is thickness of a sample.
The relative moisture uptake, $\Phi$, is defined as
\begin{equation}
\Phi(\bar{t})=\frac{R(\bar{t})}{R(\infty)},
\end{equation}
where $R(\bar{t})$ is the mass gain at the reduced time $\bar{t}$
and $R(\infty)$ is the maximal mass gain.
Observations on samples with various thicknesses (not presented)
show that the mass gain curves plotted in the reduced
coordinates $(\Phi,\bar{t})$ are superposed fairly well.

Figures 6 to 10 demonstrate that moisture diffusion in the neat resin
is roughly Fickian, but it becomes strongly anomalous with 
an increase in the filler content.

The aim of the remaining part of this work is to develop constitutive
equations for the elastoplastic behavior and moisture diffusion in
a nanocomposite and to find adjustable parameters in these relations
by matching the experimental data plotted in Figures 1 to 10.

\section{A model for the elastoplastic response}

To simplify the analysis, we model a nanocomposite as an equivalent 
network of macromolecules bridged by junctions (entanglements,
physical cross-links and particles of filler).
Because the viscoelastic behavior of the nanocomposite is disregarded, 
the network is treated as permanent (we suppose that active chains 
do not separate from junctions and dangling chains do not merge with
the netwotk within the experimental time-scale).
To describe the elastoplastic behavior of the network, we assume that
junctions slide with respect to their reference positions in the bulk
material.
For uniaxial deformation, sliding of junctions is determined by a plastic
strain $\epsilon_{\rm p}$.
Adopting the conventional hypothesis that the macro-strain, $\epsilon$,
is transmitted to all chains in the network by surrounding macromolecules, 
we write
\begin{equation}
\epsilon=\epsilon_{\rm e}+\epsilon_{\rm p},
\end{equation}
where $\epsilon_{\rm e}$ is an elastic strain.
We accept the mean-field approach and treat $\epsilon_{\rm e}$ and 
$\epsilon_{\rm p}$ as average elastic and plastic strains per chain.

It is assumed that under active loading,
the rate of changes in the strain, $\epsilon_{\rm p}$, with time, $t$,
is proportional to the rate of changes in the macro-strain $\epsilon$,
\begin{equation}
\frac{d\epsilon_{\rm p}}{dt} (t)=\varphi (\epsilon_{\rm e}(t)) 
\frac{d\epsilon}{dt}(t),
\end{equation}
where the coefficient of proportionality, $\varphi$, is a function 
of the elastic strain $\epsilon_{\rm e}$.

The following restrictions are imposed on the function $\varphi(\epsilon_{\rm e})$:
\begin{enumerate}
\item
The coefficient of proportionality in Eq. (4)  vanishes at the zero elastic strain,
$\varphi(0) =0$.

\item
The function $\varphi$ monotonically increases with $\epsilon_{\rm e}$ 
and tends to some constant $a\in (0,1)$ for relatively large elastic strains,
\[
\lim_{\epsilon_{\rm e}\to \infty} \varphi(\epsilon_{\rm e}) =a.
\]
The constant $a$ determines the rate of sliding in junctions 
at the stage of a developed plastic flow.
\end{enumerate}
To approximate experimental data, we use the function
\begin{equation}
\varphi(\epsilon_{\rm e})=a \Bigl [ 1-\exp\Bigl (-\frac{\epsilon_{\rm e}}{e}\Bigr )
\Bigr ],
\end{equation}
which satisfies the above conditions.
This function is determined by two adjustable parameters, $a$ and $e$,
where $a$ is the limiting value of $\varphi$ at ``large" elastic strains,
and $e>0$ is a constant that characterizes how ``large" 
an elastic strain, $\epsilon_{\rm e}$, is.

A chain is modelled as a linear elastic solid with the mechanical energy
\[ 
w=\frac{1}{2}\mu \epsilon_{\rm e}^{2},
\]
where $\mu$ is an average rigidity per chain.
Multiplying the energy, $w$, by the number of chains per unit volume, $n_{0}$,
we find the strain energy density 
\begin{equation}
W=\frac{1}{2}E\epsilon_{\rm e}^{2},
\end{equation}
where $E=\mu n_{0}$ is an elastic modulus.
It is worth noting that the elastic modulus, $E$, may differ from the Young modulus.
The latter is conventionally defined as the tangent of the angle between the
tangent straightline to the stress--strain curve at small strains
and the horizontal axis.
These two quantities coincide when the elastic strain, $\epsilon_{\rm e}$,
is equal to the macro-strain, $\epsilon$.
For the elasto-plastic response of a nanocomposite, when the stress--strain
diagram substantially differs from a straightline, $E$ exceeds the Young
modulus determined by the traditional method.

For isothermal uniaxial deformation, the Clausius-Duhem inequality reads,
\[
Q(t)=-\frac{dW}{dt}(t)+\sigma(t)\frac{d\epsilon}{dt}(t) \geq 0,
\]
where $Q$ is internal dissipation per unit volume.
Substition of Eqs. (3), (4) and (6) into this formula results in
\[
Q(t)= \Bigl [ \sigma(t)-E\epsilon_{\rm e}(t)
\Bigl (1-\varphi(\epsilon_{\rm e}(t))\Bigr ) \Bigr ] \frac{d\epsilon}{dt}(t)\geq 0.
\]
Assuming the expression in square brackets to vanish,
we arrive at the stress--strain relation
\begin{equation}
\sigma(t) =E\epsilon_{\rm e}(t) \Bigl [1-\varphi(\epsilon_{\rm e}(t))\Bigr ].
\end{equation}
Constitutive equations (3) to (5) and (7) are determined by 3 adjustable parameters:
\begin{enumerate}
\item
the elastic modulus $E$,

\item
the rate of developed plastic flow $a$,

\item
the strain, $e$, that characterizes transition to the steady-state
plastic deformation.
\end{enumerate}
To find these quantities, we match experimental data depicted in
Figures 1 to 5.

\section{Fitting observations in mechanical tests}

It follows from Eqs. (3), (4), (5) and (7) that in a uniaxial tensile test,
the stress is given by
\begin{equation}
\sigma(\epsilon) = E(\epsilon-\epsilon_{\rm p})\Bigl \{ 1- a 
\Bigl [1-\exp \Bigl (-\frac{\epsilon-\epsilon_{\rm p}}{e}\Bigr )
\Bigr ]\Bigr \},
\end{equation}
where the plastic strain, $\epsilon_{\rm p}$, satisfies the nonlinear 
differential equation
\begin{equation}
\frac{d\epsilon_{\rm p}}{d\epsilon}(\epsilon)
=a \Bigl [1-\exp \Bigl (-\frac{\epsilon-\epsilon_{\rm p}}{e}\Bigr )
\Bigr ]
\end{equation}
with the initial condition $\epsilon_{\rm p}(0)=0$.

We begin with matching the experimental data in a test on the neat
vinyl ester resin,  see Figure 1.
To find the  material constants, $E$, $a$ and $e$,
we fix some intervals $[0,a_{\max}]$ and $[0,e_{\max}]$, 
where the ``best-fit" parameters $a$ and $e$ are assumed to be located,
and divide these intervals into $I$ subintervals by
the points $a_{i}=i\Delta a$ and $e_{j}=j\Delta e$  ($i,j=1,\ldots,I$)
with $\Delta a=a_{\max}/I$ and $\Delta e=e_{\max}/I$.
For any pair, $\{ a_{i}, e_{j} \}$, we integrate Eq. (9) numerically
(with the step $\Delta \epsilon=1.0\cdot 10^{-5}$)
by the Runge--Kutta method.
Given $\{ a_{i}, e_{j} \}$, the elastic modulus $E=E(i,j)$ 
is found by the least-squares technique from the condition 
of minimum of the function
\[
F(i,j)=\sum_{\epsilon_{k}} \Bigl [ \sigma_{\rm exp}(\epsilon_{k})
-\sigma_{\rm num}(\epsilon_{k}) \Bigr ]^{2},
\]
where the sum is calculated over all experimental points,
$\epsilon_{k}$, depicted in Figure 1, 
$\sigma_{\rm exp}$ is the stress measured in a tensile test, 
and $\sigma_{\rm num}$ is given by Eq. (8).
The ``best-fit" parameters $a$ and $e$ minimize the function
$F$ on the set $ \{ a_{i}, e_{j} \quad (i,j=1,\ldots, I)  \}$.
Fitting the observations results in $e=2.11\cdot 10^{-3}$.

We fix this value of $e$ and proceed with matching observations in
tensile tests on other specimens by using only
two material constants, $E$ and $a$, which are determined by the 
above algorithm.
The average values of $E$ and $a$ (over 4 samples) are depicted
in Figures 11 and 12, where the vertical bars stand for the standard
deviations of these parameters.

Afterwards, we match experimental data in tensile tests on
specimens with various contents of filler, $\nu$.
The same algorithm of fitting is employed with the fixed value
$e=2.11\cdot 10^{-3}$.
Figures 1 to 5 demonstrate fair agreement between the observations
and the results of numerical simulation.
The elastic modulus, $E$, and the rate of developed plastic flow,
$a$, are plotted versus the filler content, $\nu$, in Figures 11
and 12.
The experimental data are approximated by the phenomenological
equations
\begin{equation}
E=E_{0}-E_{1}\nu,
\qquad
a=a_{0}-a_{1}\nu,
\end{equation}
where the coefficions $E_{j}$ and $a_{j}$ ($j=0,1$) are found by
the least-squares technique.

\section{A model for the moisture diffusion}

A sample is treated as a rectilinear plate with thickness $2l$.
We introduce Cartesian coordinates $\{ x,y,z \}$,
where the axis $x$ is perpendicular to the middle plane of the plate,
and the axes $y$ and $z$ lie in the middle plane.
Length and width of the plate are assumed to substantially exceed 
its thickness, which implies that the moisture concentration depends
on the coordinate $x$ only.

After the plate is immersed into water, three processes occur in the
composite material:
\begin{enumerate}
\item
sorption of water molecules to the plate faces from the surrounding,

\item
diffusion of the penetrant into the plate,

\item
adsorption of water molecules on the surfaces of clay layers,
where these molecules become immobilized.
\end{enumerate}

It is conventionally accepted that the rate of sorption in glassy polymers
noticeably exceeds the rate of diffusion (Chen et al., 2001), which implies 
that the sorption equilibrium is rapidly established.
The equilibrium condition is given by
\begin{equation}
n(t,x)\Bigl |_{x=\pm l}=n^{\circ},
\end{equation}
where $n$ is the moisture concentration at time $t$ at point $x$
[the number of water molecules in the polymeric matrix occupying
a volume with the unit area in the plane $(y,z)$ and the thickness 
$dx$ reads $n(t,x)dx$ ],
and $n^{\circ}$ is the equilibrium moisture concentration in the matrix
on the faces of the plate.

Diffusion of the penetrant through the matrix is described by 
the mass conservation law
\begin{equation}
\frac{\partial n}{\partial t}=\frac{\partial J}{\partial x}-\frac{\partial n_{1}}{\partial t},
\end{equation}
where $J(t,x)$ stands for the mass flux, 
and $n_{1}(t,x)$ denotes the concentration of water molecules immibilized
at the surfaces of clay layers.

The mass flux obeys the Fick equation
\begin{equation}
J=D\frac{\partial n}{\partial x},
\end{equation}
where $D$ is diffusivity.
Equation (13) is tantamount to the assumption that diffusion occurs
through the polymeric matrix only.
However, the coefficient of diffusion, $D$, is assumed to depend
on the volume content of filler, because the presence of
particles results in a decay in the molecular mobility of
polymeric molecules which provides the driving force for 
diffusion of penetrants.

Adsorption of water molecules on the surfaces of filler is determined
by the first-order kinetic equation
\begin{equation}
\frac{\partial n_{1}}{\partial t}=kn(n_{1}^{\circ}-n_{1}).
\end{equation}
According to Eq. (14), the rate of adsorption is proportional to the
concentration of the penetrant in the matrix, $n$, and to the current
number of ``unoccupied sites" on the surfaces of filler, 
$n_{1}^{\circ}-n_{1}$, where $n_{1}^{\circ}$ is the total number of
sites where water molecules can be immobilized.
In the general case, the rate of adsorption, $k$, and the maximal
concentration of unoccupied sites, $n_{1}^{\circ}$,
are functions of the clay content, because the presence of nanoparticles
affect mobility of chains in the polymeric matrix, and, as a
consequence, their chemical potential.

Neglecting the moisture content in a sample before testing, we adopt
the following initial conditions:
\begin{equation}
n(t,x)\Bigl |_{t=0}=0,
\qquad
n_{1}(t,x)\Bigl |_{t=0}=0.
\end{equation}
Equations (12) to (14) with initial conditions (15) and boundary condition (11)
uniquelly determine the diffusion process.
The moisture mass gain (per unit mass of the sample) is given by
\begin{equation}
R(t) =\frac{\kappa}{2l\rho} \int_{-l}^{l} \Bigl [ n(t,x)+n_{1}(t,x) \Bigl ] dx,
\end{equation}
where $\kappa$ is the average mass of a penetrant molecule,
and $\rho$ is mass density of the matrix.

Our aim now is to transform Eqs. (11) to (16).
For this purpose, we introduce the dimensionless coordinate, $\bar{x}=x/l$,
and the reduced moisture concentrations, 
$c=n/n^{\circ}$ and $c_{1}=n_{1}/n^{\circ}$.
Combining Eqs. (1), (12) and (13), we arrive at the mass flux equation
\begin{equation}
\frac{\partial c}{\partial \bar{t}}=\bar{D}\frac{\partial^{2}c}{\partial \bar{x}^{2}}
-\frac{\partial c}{\partial \bar{t}},
\end{equation}
where 
\begin{equation}
\bar{D}=4D.
\end{equation}
In the new notation, Eq. (14) reads
\begin{equation}
\frac{\partial c_{1}}{\partial \bar{t}}=Kc(C-c_{1}),
\end{equation}
where 
\begin{equation}
K=4kl^{2}n^{\circ},
\qquad
C=\frac{n_{1}^{\circ}}{n^{\circ}}.
\end{equation}
Initial conditions (15) remain unchanged,
\begin{equation}
c(0,\bar{x})=0,
\qquad
c_{1}(0,\bar{x})=0.
\end{equation}
Bearing in mind that the functions $c(\bar{t},\bar{x})$ and $c_{1}(\bar{t},\bar{x})$ 
are even functions of $\bar{x}$, we present boundary condition  (11) in the
form
\begin{equation}
\frac{\partial c}{\partial \bar{x}}(\bar{t},0)=0,
\qquad
c(\bar{t},1)=1.
\end{equation}
Finally, Eq. (16) is given by
\begin{equation}
R(\bar{t})=\frac{\kappa}{\rho}  n^{\circ} \int_{0}^{1} \Bigl [ c(\bar{t},\bar{x})+c_{1}(\bar{t},\bar{x})
\Bigr ] d\bar{x}.
\end{equation}
With an increase in time, $\bar{t}$, the solutions of Eqs. (17) and (19) tend
to the steady-state solutions
\begin{equation}
c(\infty,\bar{x})=1,
\qquad
c_{1}(\infty,\bar{x})=C.
\end{equation}
It follows from Eqs. (23) and (24) that the maximal moisture uptake reads
\begin{equation}
R(\infty)=\frac{\kappa}{\rho} n^{\circ}(1+C),
\end{equation}
whereas the relative mass gain, see Eq. (2), is given by
\begin{equation}
\Phi(\bar{t})=\frac{1}{1+C}\int_{0}^{1} \Bigl [ c(\bar{t},\bar{x})+c_{1}(\bar{t},\bar{x})
\Bigr ] d\bar{x}.
\end{equation}
Equations (17), (19) and (26) with initial conditions (21) and boundary conditions (22)
are determined by 3 adjustable parameters:
\begin{enumerate}
\item
the reduced diffusivity $\bar{D}$,

\item
the rate of moisture adsorption on nanoparticles $K$,

\item
the reduced concentration of unoccupied sites on the surfaces
of filler $C$.
\end{enumerate}
These constants are found by fitting the experimental data in diffusion tests.

\section{Fitting observations in diffusion tests}

We begin with matching the experimental data for the neat vinyl resin
depicted in Figure 6.
Because in the absence of filler, moisture is not adsorbed on the surfaces
of particles, we set $C=0$ in Eq. (17).
Equations (17), (21) and (22) are reduced to the conventional boundary problem
for the diffusion equation
\begin{equation}
\frac{\partial c}{\partial \bar{t}}=\bar{D}\frac{\partial^{2}c}{\partial \bar{x}^{2}},
\qquad
\frac{\partial c}{\partial \bar{x}}(\bar{t},0)=0,
\qquad
c(\bar{t},1)=1,
\qquad
c(0,\bar{x})=0,
\end{equation}
which is resolved by the finite-difference method with an explicit algorithm.
Given a coefficient of diffusion, $\bar{D}$, we divide the interval $[0,1]$
into $M$ subintervals by points $\bar{x}_{m}=m\Delta x$  $(m=0,1,\ldots,M)$
with $\Delta x=1/M$, introduce discrete time $\bar{t}_{n}=n\Delta t$,
and replace Eq. (27) by its finite-difference approximation
\begin{eqnarray}
c(0,\bar{x}_{m}) &=& 0
\quad
(m=0,1,\ldots,M-1),
\qquad
c(0,\bar{x}_{M})=1,
\nonumber\\
c(\bar{t}_{n+1},\bar{x}_{m}) &=& c(\bar{t}_{n},\bar{x}_{m})+\bar{D}\frac{\Delta t}{\Delta x^{2}}
\Bigl [ c(\bar{t}_{n},\bar{x}_{m+1})-2c(\bar{t}_{n},\bar{x}_{m})
\nonumber\\
&& +c(\bar{t}_{n},\bar{x}_{m-1}) \Bigr ]
\quad
(m=1,\ldots,M-1),
\nonumber\\
c(\bar{t}_{n+1},\bar{x}_{0}) &=& c(\bar{t}_{n+1},\bar{x}_{1}),
\qquad
c(\bar{t}_{n+1},\bar{x}_{M})=1
\qquad
(n=1,2,\ldots).
\end{eqnarray}
To calculate the relative mass gain, the integral in Eq. (26) is approximated by
the Euler formula
\begin{equation}
\Phi(\bar{t}_{n})=\Delta x\sum_{m=0}^{M-1} c(\bar{t}_{n},\bar{x}_{m}).
\end{equation}

The coefficient $\bar{D}$ is determined by the following procedure.
We fix some interval $[D_{\min},D_{\max}]$, where the ``best-fit" value
of $\bar{D}$ is supposed to be located, and divide this interval
into $I$ sub-intervals by the points $\bar{D}_{i}=D_{\min}+i\Delta D$
($i=0,1,\ldots,I$), where $\Delta D=(D_{\max}-D_{\min})/I$.
For any $\bar{D}_{i}$, Eq. (28) is solved numerically (with the steps
$\Delta x=0.05$ and $\Delta t=0.005$ that ensure the stability
of the numerical algorithm), and the function $\Phi(\bar{t}_{n})$ is
given by Eq. (29).
The ``best-fit" value of $\bar{D}$ is found from the condition
of minimum of the function
\[
F=\sum_{\bar{t}_{k}} \Bigl [ \Phi_{\exp}(\bar{t}_{k})-\Phi_{\rm num}(\bar{t}_{k})
\Bigr ]^{2},
\]
where the sum is calculated over all point $\bar{t}_{k}$ depicted in Figure 6,
$\Phi_{\exp}(\bar{t}_{k})$ is the relative mass gain measured in the test,
and $\Phi_{\rm num}(\bar{t}_{k})$ is determined from Eq. (29).
After the ``best-fit" diffusivity, $\bar{D}_{i}$, is found, the procedure is
repeated for the new interval $[\bar{D}_{i-1},\bar{D}_{i+1}]$ to guarantee
an acceptable accuracy of matching observations.
Figure 6 demonstrates fair agreement between the experimental data
and the results of numerical simulation.

To ensure that the numerical results are independent of the choice of
the steps of integration, $\Delta x$ and $\Delta t$, we repeat integration
of Eqs. (28) and (29) with the ``best-fit" parameter $\bar{D}$ and the new
values $\Delta x=0.01$ and $\Delta t=0.0005$.
No difference was observed between the curve plotted in Figure 6 and
the curve obtained by the numerical integration of Eqs. (28) and (29)
with the decreased steps in time and the spatial coordinate.

The above procedure is repeated to match observations in the tests on
4 different samples.
The average value of diffusivity is presented in Figure 13, where the vertical
bars stand for the standard deviation.

It is worth noting that the algorithm of fitting experimental data used
in this work differs from the conventional procedure employed by 
Shah et al. (2002), according to which the coefficient of diffusion 
was determined by matching the initial parts of the mass gain diagrams 
(where the function $\Phi(\bar{t})$ is approximately linear).

The same numerical algorithm is employed to match experimental data
on polymeric nanocomposites with various amounts of filler.
The difference in the treatment of observations consists in the following:
(i) Eq. (28) are replaced by the finite-difference approximation of
Eqs. (17) and (19);
(ii) the ``best-fit" values are simultaneously determined for 3 adjustable
parameters: $\bar{D}$, $C$ and $K$.
Figures 7 to 10 show good agreement between the experimental data 
and the results of numerical analysis.
The average values of $D$ [this quantity is found from Eq. (18)], 
$C$ and $K$ (over 4 tests) are plotted versus the filler fraction, $\nu$, 
in Figures 13 to 15.
The dependences of these parameters on the concentration of filler
are approximated by the linear equations
\begin{equation}
D=D_{0}-D_{1}\nu,
\qquad
C=C_{1}\nu,
\qquad
K=K_{0}-K_{1}\nu,
\end{equation}
where the coefficions $D_{j}$, $C_{j}$ and $K_{j}$ ($j=0,1$) are found by
the least-squares technique.

It follows from Eq. (25) that the maximal moisture uptake per unit mass of the matrix,
$R_{\rm pol}=\kappa n^{\circ}/\rho$, is given by
\begin{equation}
R_{\rm pol}=\frac{R(\infty)}{1+C}.
\end{equation}
Using experimental data for $R(\infty)$ and the values of $C$ determined by
matching observations, we calculate $R_{\rm pol}$ from Eq. (31) and plot 
this parameter versus the clay content in Figure 16.

\section{Discussion}

Figure 11 demonstrates that the elastic modulus, $E$, decreases with
the clay content, $\nu$, in the region of small concentrations of filler
($\nu<1.0$ wt.-\%), and remains practically constant when $\nu$
exceeds 1.0 wt.-\%.
According to Figure 16, the filler fraction, $\nu$, affects the maximal 
moisture uptake, $R_{\rm pol}$, in a similar way.
These dependences of the parameters $E$ and $R_{\rm pol}$
seem quite natural, because these quantities characterize the 
macro-response of a nanocomposite in tensile and diffusion tests.
They may be explained by a progressive decrease in the degree
of cross-linking with the growth of clay concentration
(Bharadwaj et al., 2002), which results in a decrease in
the number of active chains (whose ends are linked to nearby
junctions) compared to the neat vinyl ester resin.
This implies that the elastic modulus (proportional to the number of active
chains) is reduced and the maximal weight gain (proportional to
the free volume per unit active chains) is increased.
Figures 11 and 16 show that the concentration of chains 
reaches its limiting value rather rapidly (when the filler content 
is about $\nu_{\rm c}=1$ wt.-\%), and an increase in the clay concentration 
above this critical value does not affect the cross-linking process.

Figure 12 demonstrates that the rate of developed plastic flow,
$a$, strongly decreases with $\nu$ in the region below the
critical concentration of filler, $\nu\leq \nu_{\rm c}$,
and it weakly decreases with $\nu$ when the clay content 
exceeds its critical value.
Figures 13 and 15 reveal that the coefficient of diffusion, $D$, 
and the rate of adsorption of water molecules on surfaces of 
nano-layers, $K$, demonstrate similar dependences on 
the filler fraction.
This behavior of the quantities $a$, $D$ and $K$ appears 
to be in agreement with our physical intuition, because
all these parameters reflect mobility of polymeric chains.
According to Figures 12, 13 and 15, this mobility is
drastically reduced with $\nu$ in the sub-critical region,
and it proceeds to decrease (however, rather weakly) when the
filler content exceeds the critical concentration $\nu_{\rm c}$.
The conclusion that the mobility of chains in the polymeric matrix
decreases with the clay content is also confirmed by DSC
measurements (Shah et al., 2002) which reveal the the glass
transition temperature monotonically increases with $\nu$.

Figure 14 reveals that the parameter $C$ linearly increases with $\nu$.
It follows from Figure 16 and Eqs. (25) and (31) that the maximal moisture
content in the polymeric matrix, $n^{\circ}$, is independent of the
clay fraction.
This result together with Eq. (20) and Figure 14 implies that 
number of sites on the clay layers where water molecules 
are immobilized, $n_{1}^{\circ}$, linearly grows with the filler
concentration.
The latter assertion is in agreement with the main assumptions of
our model (the total number of sites where the penetrant molecules 
are immobilized equals the product of the average number of these
sites per particle by the number of clay particles), and it
may be treated as a confirmation of the kinetic equations for the anomalous
moisture diffusion.

The similarity in this behavior of the quantities $a$, $D$ and $K$ 
as functions of the clay concentration, $\nu$, provides a way to
substantially reduce the number of (time-consuming) moisture
diffusion experiments.
If the parameters $a$ and $D$ decrease with $\nu$ in the same fashion
(our analysis of observations on vinyl ester resin nanocomposite
with MMT particles treated with vinyl benzine trimethyl ammonium
chloride confirms this hypothesis), the function $a(\nu)$ can be 
determined with a high level of accuracy in standard tensile tests, 
whereas only a few moisture diffusion tests are necessary to 
find the dependence of diffusivity on the clay content, $D=D(\nu)$.

\section{Concluding remarks}

Mechanical and diffusion tests have been performed
on a nanocomposite with vinyl ester resin matrix and montmorillonite
clay filler at room temperature.
Experimental data in tensile tests demonstrate that the mechanical
behavior of the neat vinyl ester resin is strongly elastoplastic,
whereas an increase in the clay content results in a decrease
in the plastic strain (due to the constrains on mobility of junctions
between polymeric chains imposed by nano-particles).
Observations in diffusion tests show that moisture diffusion in
the neat resin is Fickian, whereas it becomes noticeably 
anomalous (non-Fickian) with the growth of the clay content.
This transition is ascribed by adsorbtion (immobilization) 
of the penetrant molecules on surfaces of hydrophilic clay layers.

Constitutive models have been proposed to describe the elastoplastic
response and anomalous moisture diffusion in a nanocomposite.
Both models are determined by 3 adjustable parameters that
are found by fitting the experimental data.
Fair agreement is demonstrated between the observations and
the results of numerical simulation.

The following conclusions are drawn:
\begin{enumerate}
\item
A critical concentration of clay is found, $\nu_{\rm c}\approx 1$ wt.-\%,
corresponding to both mechanical and diffusion tests. 
The filler content, $\nu$, strongly affects the material constants
in the sub-critical region, and its influence becomes rather weak
in the post-critical domain.

\item
The quantities independent of molecular mobility, such as
the elastic modulus, $E$, and the maximal moisture uptake by the
polymeric matrix, $R_{\rm pol}$, are not affected by the concentration
of clay particles when the filler content exceeds its critical value.

\item
The quantities reflecting mobility of polymeric chains,
such as the rate of developed plastic flow, $a$,
the coefficient of diffusion, $D$,
and the rate of moisture adsorption on nano-particles, $k$,
strongly decrease with the filler content in the sub-critical
region, and proceed to decrease (with smaller rates)
at $\nu>\nu_{\rm c}$.
\end{enumerate}
\newpage
\section*{References}
\parindent 0mm

Bavisi BH, Pritchard G, Ghotra JS (1996)
Measuring and reducing moisture penetration 
through thick laminates.
Adv. Polym. Technol. 15: 223--235
\vspace*{1 mm}

Bharadwaj RK, Mehrabi AR, Hamilton C, Trujillo C, Murga M, 
Fan R, Chavira A, Thompson AK (2002)
Structure--property relationships in cross-linked polyester--clay
na\-no\-composites.
Polymer 43: 3699--3705
\vspace*{1 mm}

Boyce MC, Parks DM, Argon AS (1988)
Large inelastic deformation of glassy polymers.
1. Rate dependent constitutive model.
Mech. Mater. 7: 15--33
\vspace*{1 mm}

Carter HG, Kibler KG (1978)
Langmuir-type model for anomalous diffusion in composite resin.
J. Compos. Mater. 12: 118--130
\vspace*{1 mm}

Cai LW, Weitsman Y (1994)
Non-Fickian moisture diffusion in polymeric composites.
J. Compos. Mater. 28: 130--154
\vspace*{1 mm}

Chang Y-W, Yang Y, Ryu S, Nah C (2002)
Preparation and properties of EPDM/organo\-mon\-tmorillonite hybrid
nanocomposites.
Polym. Int. 51: 310--324
\vspace*{1 mm}

Chen C, Han B, Li J, Shang T, Zou J, Jiang W (2001)
A new model on the diffusion of small molecule penetrants in
dense polymer membrances.
J. Membrane Sci. 187: 109--118
\vspace*{-2 mm}

Coulon G, Lefebvre JM, Escaig B (1986)
The preyield evolution with strain of the work-hardening rate of
glassy polymers (PABM resin).
J. Mater. Sci. 21: 2059--2066
\vspace*{1 mm}

David L, Quinson R, Gauthier C, Perez J (1997)
The role of anelasticity in high stress mechanical response
and physical properties of glassy polymers.
Polym. Eng. Sci. 37: 1633--1640
\vspace*{1 mm}

Drozdov AD (2001)
The viscoelastic and viscoplastic responses of glassy polymers.
Int. J. Solids Structures 38: 8259--8278
\vspace*{1 mm}

Gurtin ME, Yatomi C (1979)
On a model for two phase diffusion in composite mateirals.
J. Compos. Mater. 13: 126--130
\vspace*{1 mm}

Hasan OA, Boyce MC (1995)
A constitutive model for the nonlinear viscoelastic viscoplastic
behavior of glassy polymers.
Polym. Eng. Sci. 35: 331--344
\vspace*{1 mm}

Kim TH, Jang LW, Lee DC, Choi HJ, Jhon MS (2002)
Synthesis and rheology of intercalated 
polystyrene/Na$^{+}$-montmorillonite nanocomposites.
Macromol. Rapid Commun. 23: 191--195
\vspace*{1 mm}

Ma J, Qi Z, Hu Y (2001)
Synthesis and characterization of polypropylene/clay nanocomposites.
J. Appl. Polym. Sci. 82: 3611--3617
\vspace*{1 mm}

Mangion MBM, Cavaille JY, Perez J (1992)
A molecular theory for the sub--$T_{\rm g}$ plastic mechanical
response of amorphous polymers.
Phil. Magazine A 66: 773--796
\vspace*{1 mm}

Marcovich NE, Reboedo MM, Aranguren MI (1999)
Moisture diffusion in polyester--woodflour composites.
Polymer 40: 7313--7320
\vspace*{1 mm}

Masenelli-Varlot K, Reynard E, Vigier G, Varlet J (2002)
Mechanical properties of clay-reinforced polyamide.
J. Polym. Sci. Part B: Polym. Phys. 40: 272--283
\vspace*{1 mm}

Metzler R, Klafter J (2000)
The random walk's guide to anomalous diffusion:
a fractional dynamics approach.
Phys. Rep. 339: 1--77
\vspace*{1 mm}

Patel SV, Raval DK, Thakkar JR (1999)
Novel vinyl ester resin and its urethane derivatives 
for glass reinforced composites.
Angew. Makromol. Chem. 265: 13--15
\vspace*{1 mm}

Roy S, Xu, WX, Park SJ, Liechti KM (2000)
Anomalous moisture diffusion in viscoelastic polymers:
modeling and testing.
Trans. ASME. J. Appl. Mech. 67: 391--396
\vspace*{1 mm}

Roy S, Xu W, Patel S, Case S (2001)
Modeling of moisture diffusion in the presence of bi-axial damage
in polymer matrix composite laminates.
Int. J. Solids Structures 38: 7627--7641
\vspace*{1 mm}

Shah AP, Gupta RK, GangaRao HVS, Powell CE (2002)
Moisture diffusion through vinyl ester nanocomposites made with
montmorillonite clay.
Polym. Eng. Sci. (accepted)
\vspace*{1 mm}

Tyan H-L, Wei K-H, Hsieh T-E (2000)
Mechanical properties of clay-polyimide (BTDA-ODA) nanocomposites via
ODA-modified organiclay.
J. Polym. Sci. Part B: Polym. Phys. 38: 2873--2878
\vspace*{1 mm}

Valea A, Martinez I, Gonzalez ML, Eceiza A, Mondragon I (1998)
Influence of cure schedule and solvent exposure on 
the dynamic mechanical behavior of a vinyl ester resin 
containing glass fibers.
J. Appl. Polym. Sci. 70: 2595--2602
\vspace*{1 mm}

Vanlandingham MR, Eduljee RF, Gillespie JW (1999)
Moisture diffusion in epoxy systems.
J. Appl. Polym. Sci. 71: 787--798
\vspace*{1 mm}

Uschitsky M, Suhir E (2001)
Moisture diffusion in epoxy molding compounds filled with particles.
Trans. ASME. J. Electronic Packaging
123: 47--51
\newpage

\setlength{\unitlength}{1.0 mm}
\begin{figure}[tbh]
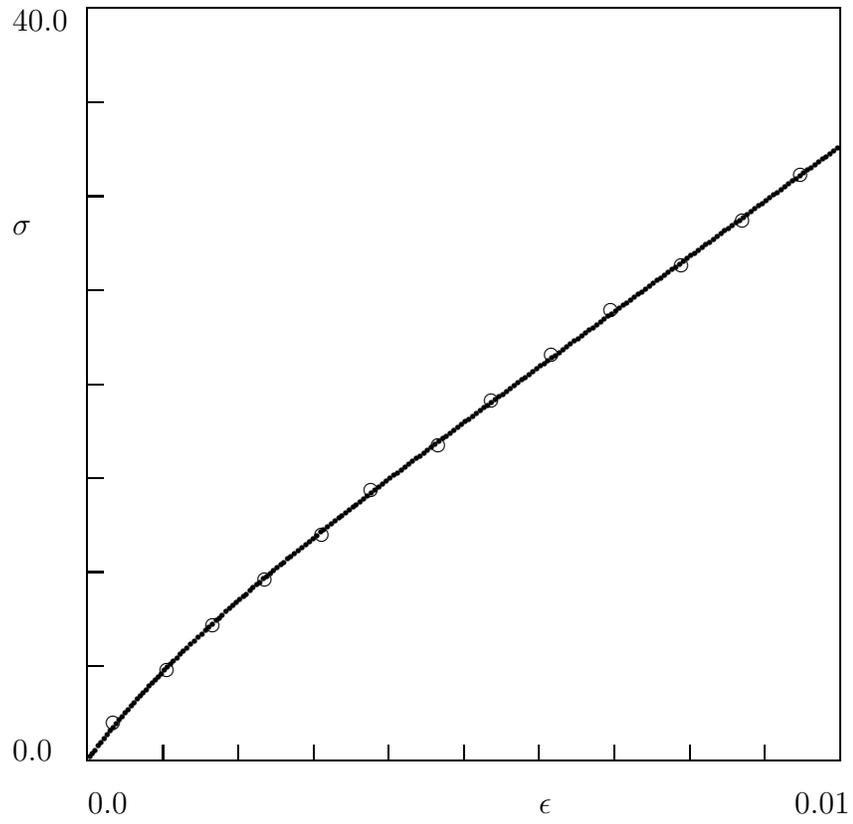

\begin{center}

\end{center}
\vspace*{10 mm}

\caption{The stress $\sigma$ MPa versus strain $\epsilon$
for neat vinyl ester resin.
Circles: experimental data.
Solid line: results of numerical simulation}
\end{figure}

\begin{figure}[tbh]
\begin{center}
%
\end{center}
\vspace*{10 mm}

\caption{The stress $\sigma$ MPa versus strain $\epsilon$
for nanocomposite with $\nu=0.5$ wt.-\%.
Circles: experimental data.
Solid line: results of numerical simulation}
\end{figure}

\begin{figure}[tbh]
\begin{center}
%
\end{center}
\vspace*{10 mm}

\caption{The stress $\sigma$ MPa versus strain $\epsilon$
for nanocomposite with $\nu=1.0$ wt.-\%.
Circles: experimental data.
Solid line: results of numerical simulation}
\end{figure}

\begin{figure}[tbh]
\begin{center}
%
\end{center}
\vspace*{10 mm}

\caption{The stress $\sigma$ MPa versus strain $\epsilon$
for nanocomposite with $\nu=2.5$ wt.-\%.
Circles: experimental data.
Solid line: results of numerical simulation}
\end{figure}

\begin{figure}[tbh]
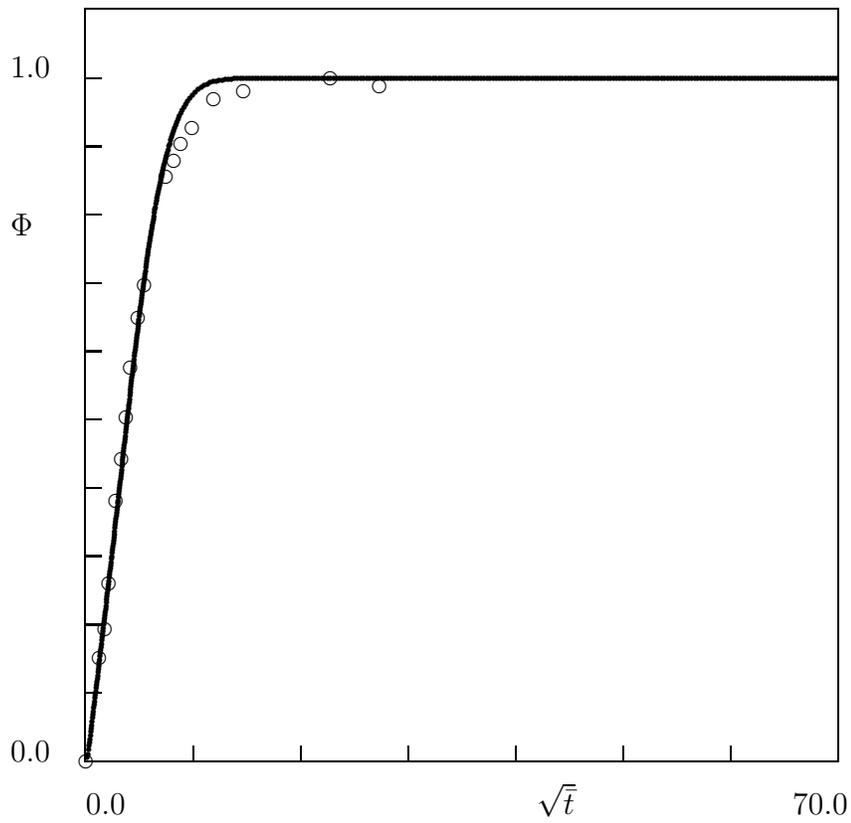

\begin{center}
%
\end{center}
\vspace*{10 mm}

\caption{The relative moisture uptake $\Phi$ versus
the reduced time $\bar{t}$ (h$^{\frac{1}{2}}$/mm) for 
neat vinyl ester resin.
Circles: experimental data ($2l=0.56726$ mm).
Solid line: results of numerical simulation}
\end{figure}

\begin{figure}[tbh]
\begin{center}
%
\end{center}
\vspace*{10 mm}

\caption{The relative moisture uptake $\Phi$ versus
the reduced time $\bar{t}$ (h$^{\frac{1}{2}}$/mm) for 
nanocomposite with $\nu=0.5$ wt.-\%.
Circles: experimental data ($2l=0.25373$ mm).
Solid line: results of numerical simulation}
\end{figure}

\begin{figure}[tbh]
\begin{center}
%
\end{center}
\vspace*{10 mm}

\caption{The relative moisture uptake $\Phi$ versus
the reduced time $\bar{t}$ (h$^{\frac{1}{2}}$/mm) for 
nanocomposite with $\nu=1.0$ wt.-\%.
Circles: experimental data ($2l=0.20108$ mm).
Solid line: results of numerical simulation}
\end{figure}

\begin{figure}[tbh]
\begin{center}
%
\end{center}
\vspace*{10 mm}

\caption{The relative moisture uptake $\Phi$ versus
the reduced time $\bar{t}$ (h$^{\frac{1}{2}}$/mm) for 
nanocomposite with $\nu=2.5$ wt.-\%.
Circles: experimental data ($2l=0.22013$ mm).
Solid line: results of numerical simulation}
\end{figure}

\begin{figure}[tbh]
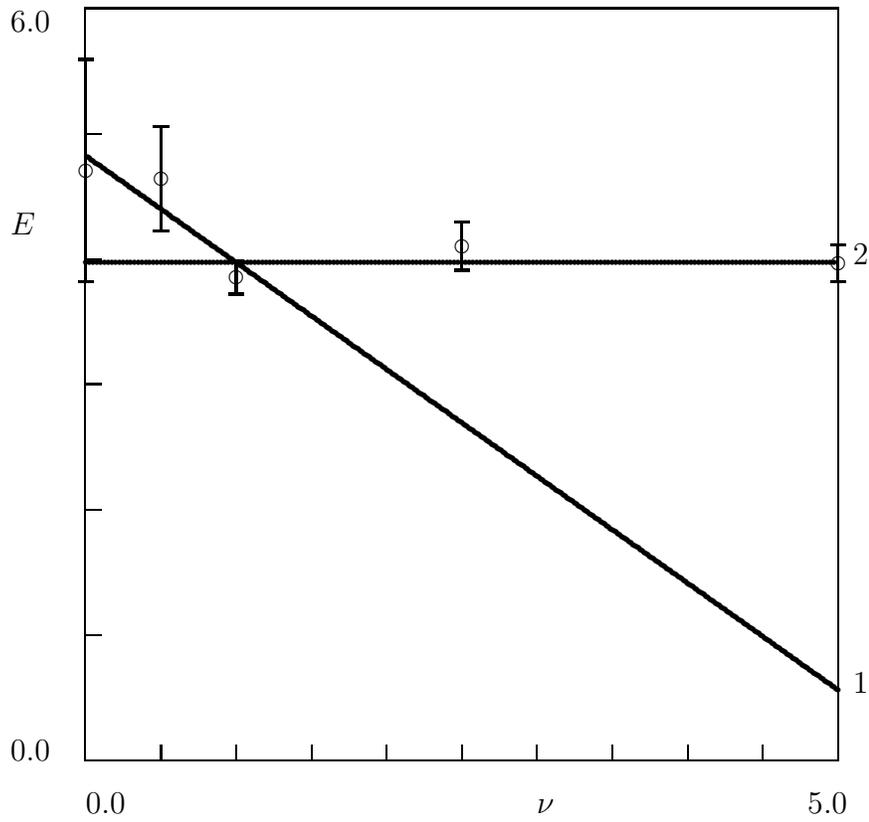

\begin{center}
%
\end{center}
\vspace*{10 mm}

\caption{The elastic modulus $E$ GPa versus the clay 
concentration $\nu$ wt.-\%.
Circles: treatment of observations.
Solid lines: approximation of the experimental data
by Eq. (10).
Curve 1: $E_{0}=4.83$, $E_{1}=0.85$;
curve 2: $E_{0}=3.97$, $E_{1}=0.0$}
\end{figure}

\begin{figure}[tbh]
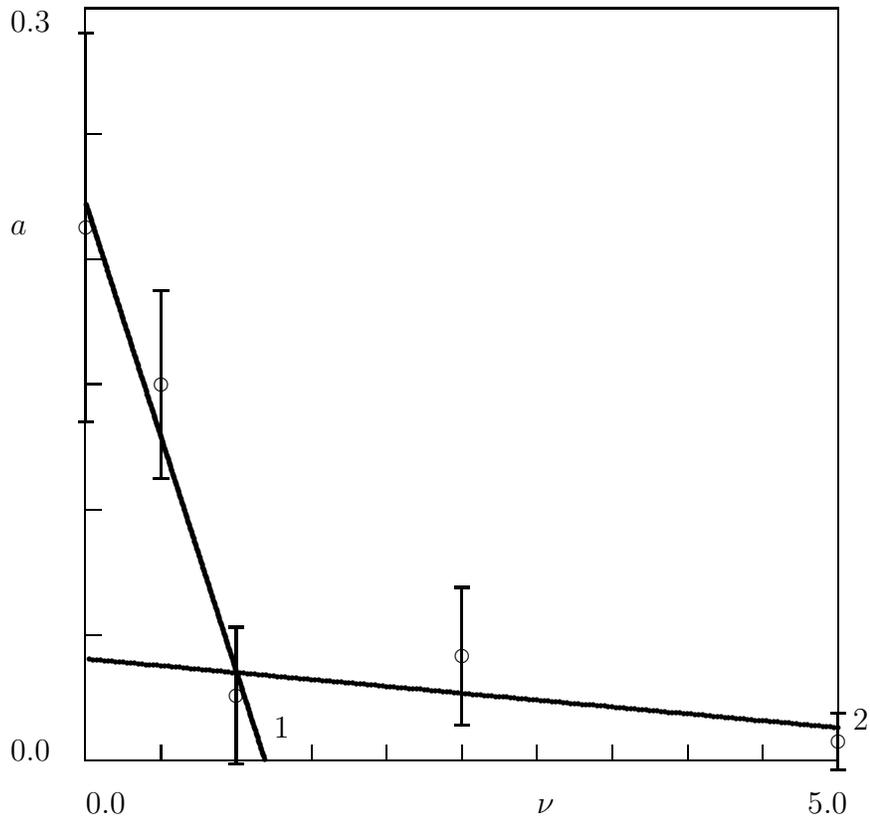

\begin{center}
%
\end{center}
\vspace*{10 mm}

\caption{The rate of developed plastic flow $a$ 
versus the clay concentration $\nu$ wt.-\%.
Circles: treatment of observations.
Solid lines: approximation of the experimental data
by Eq. (10).
Curve 1: $a_{0}=0.22$, $a_{1}=0.19$;
curve 2: $a_{0}=0.04$, $a_{1}=0.01$}
\end{figure}

\begin{figure}[tbh]
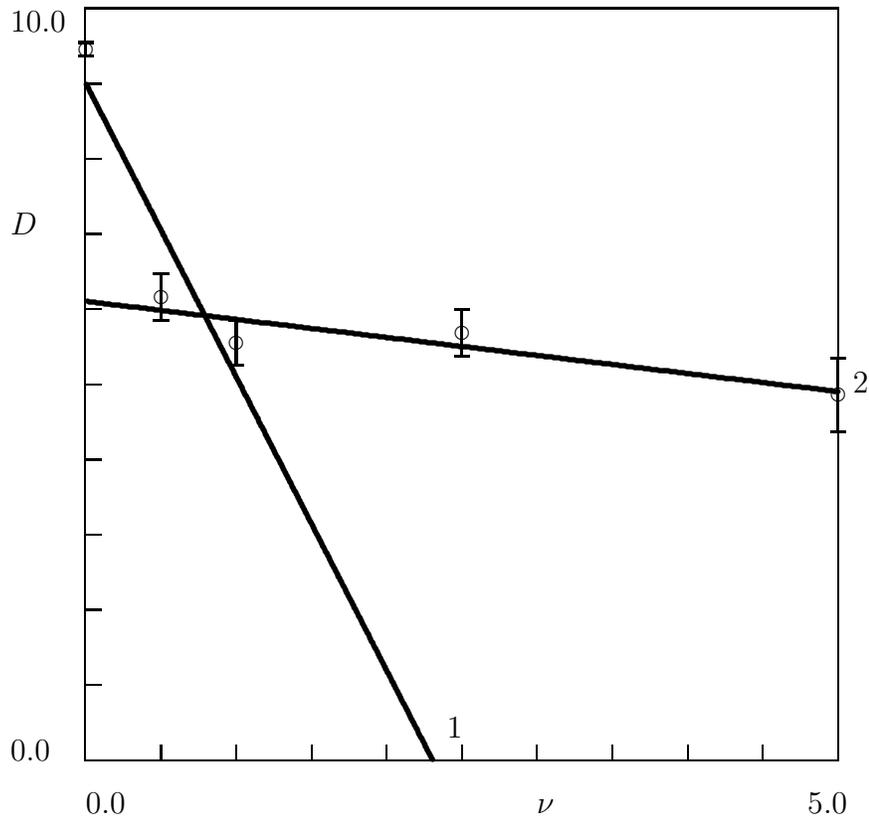

\begin{center}
%
\end{center}
\vspace*{10 mm}

\caption{Diffusivity $D\cdot 10^{-7}$ mm$^{2}$/s
versus the clay concentration $\nu$ wt.-\%.
Circles: treatment of observations.
Solid lines: approximation of the experimental data
by Eq. (31).
Curve 1: $D_{0}=9.01$, $D_{1}=3.91$;
curve 2: $D_{0}=6.11$, $D_{1}=0.24$}
\end{figure}

\begin{figure}[tbh]
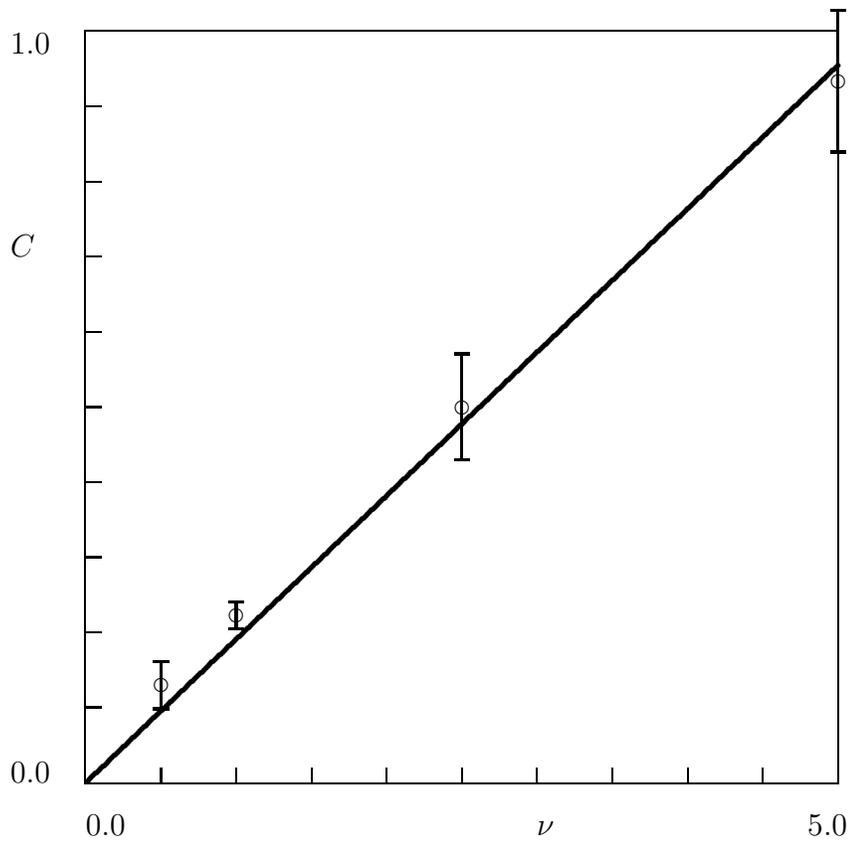

\begin{center}
%
\end{center}
\vspace*{10 mm}

\caption{The dimensionless parameter $C$ 
versus the clay concentration $\nu$ wt.-\%.
Circles: treatment of observations.
Solid line: approximation of the experimental data
by Eq. (31) with $C_{1}=0.19$}
\end{figure}

\begin{figure}[tbh]
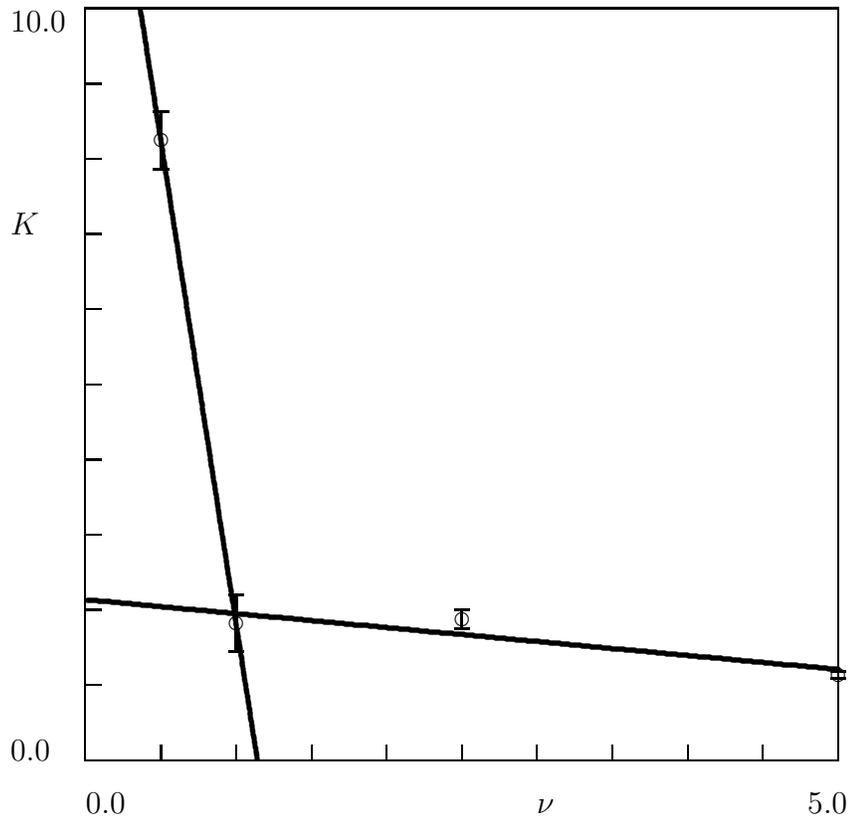

\begin{center}
%
\end{center}
\vspace*{10 mm}

\caption{The adsorption rate $K\cdot 10^{-3}$ h$^{-1}$
versus the clay concentration $\nu$ wt.-\%.
Circles: treatment of observations.
Solid lines: approximation of the experimental data
by Eq. (31).
Curve 1: $K_{0}=14.68$, $K_{1}=12.85$;
curve 2: $K_{0}=2.14$, $K_{1}=0.19$}
\end{figure}

\begin{figure}[tbh]
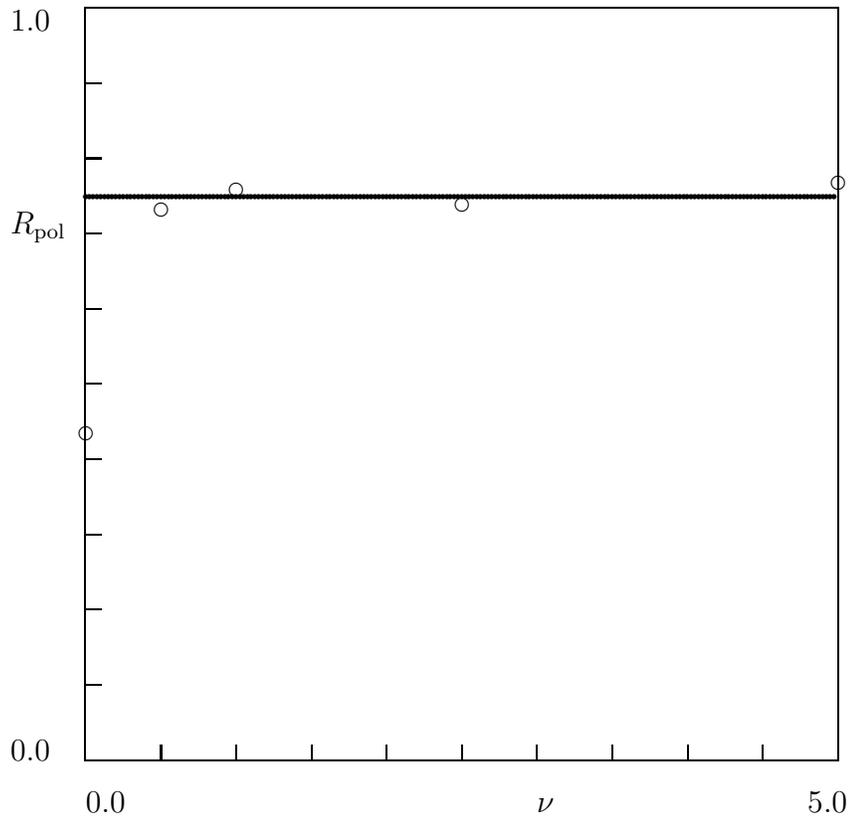

\begin{center}
%
\end{center}
\vspace*{10 mm}

\caption{The maximal moisture uptake per unit mass of the matrix
$R_{\rm pol}$ \% versus the clay concentration $\nu$ wt.-\%.
Circles: treatment of observations.
Solid line: approximation of the experimental data
by the constant $R_{\rm pol}=0.75$ \%}
\end{figure}
\end{document}